\documentclass[twocolumn,tighten,twocolappendix]{aastex631}
\usepackage{CJK}
\usepackage{amsbsy}
\usepackage{multirow}
\usepackage{bm}
\usepackage{ulem}
\usepackage{graphicx}     
\usepackage{chngcntr}     

\usepackage{epsfig}
\usepackage{color}
\usepackage{graphicx}
\usepackage{longtable}
\usepackage{hyperref}
\usepackage{amsmath}

\def\e{\,{\boldsymbol{e_1}}}

\received{XXX}
\revised{YYY}
\accepted{ZZZ}
\published{HHH}

\submitjournal{ApJL}

\counterwithout{figure}{section}

\begin{document}
\begin{CJK*}{UTF8}{gbsn}

\title{Observational Evidence for Spin Alignment Between Galaxy Groups and Their Central Galaxies}
\shorttitle{group-central spin}

\shortauthors{Peng Wang}
\correspondingauthor{Peng Wang}
\email{pwang@shao.ac.cn}

\author[0000-0003-2504-3835]{Peng Wang (王鹏)}
\affil{Shanghai Astronomical Observatory, Chinese Academy of Sciences, Nandan Road 80, Shanghai 200030, People's Republic of China.}



\begin{abstract}

We present a systematic observational study of the spin alignment between galaxy groups and their central galaxies using a large spectroscopic sample from the Sloan Digital Sky Survey. Unlike previous studies based on morphology or major axis alignment, we use spin as a direct, dynamically motivated probe. To match the limitations of observational data, the group spin, central-galaxy spin, and alignment angle are defined and measured in projection on the sky plane. By applying two novel spin estimators for galaxy groups, we find strong statistical evidence for a preferential alignment, with a mean projected angle of $34.17^\circ \pm 0.29^\circ$, significantly deviating from random expectation at $37.77\sigma$. This alignment signal persists across a wide range of group and galaxy properties, but its strength is modulated by mass, morphology, and color. Specifically, we find that more massive groups and more massive central galaxies exhibit stronger alignment. Furthermore, elliptical central galaxies show stronger alignment than spirals, and bluer central galaxies are more strongly aligned than redder ones. Our results suggest a close dynamical link between the spin of central galaxies and their host groups, modulated by their physical properties and star formation history. These results provide new insights into the dynamical connection between central galaxies and their host dark matter halos.

\end{abstract}



\keywords{
    \href{http://astrothesaurus.org/uat/597}{Galaxy groups (597)};
    \href{https://astrothesaurus.org/uat/602}{Galaxy kinematics (602)};
    \href{http://astrothesaurus.org/uat/1882}{Astrostatistics (1882)}}


\section{Introduction} 
\label{sec:intro}

The hierarchical model of structure formation posits that galaxy groups are assembled through the accretion and merging of smaller halos, acquiring their overall properties—including angular momentum—via large-scale tidal torques and merger events \citep{1984ApJ...286...38W, 1969ApJ...155..393P}. At the center of these groups typically resides \citep{2005ApJ...625..613A} the brightest group galaxy (BCG), whose formation and evolution are influenced by both the group-scale environment and the underlying dark matter halo. Theoretically, the central galaxy is expected to exhibit a certain degree of alignment with its host group, reflecting the dynamical connection established during hierarchical assembly.

Observationally, the most accessible measure of alignment has been the comparison between the major axes of galaxies and those of their host groups, which can be traced either by the spatial distribution of member galaxies \citep{2006MNRAS.369.1293Y, 2018ApJ...859..115W} or by the X-ray morphology of the intragroup medium \citep{2017NatAs...1E.157W}. Extensive observational studies have shown that the typical alignment angle between the major axes of central galaxies and their host halos generally falls in the range of $20^\circ$ to $40^\circ$. For example, \cite{2004MNRAS.347..895H} found typical alignment angles of $\sim30^\circ$ between galaxies and their dark matter halos, later constrained by \cite{2009ApJ...694..214O} to $35\pm2^\circ$. \cite{2008MNRAS.385.1511W} reported stronger alignment for more massive halos, while the alignment signal for blue galaxies was weak.

On the theoretical side, hydrodynamical simulations by \cite{2014MNRAS.441..470T,2015MNRAS.453..469T} showed that the three-dimensional mean alignment angle between the dark matter and stellar distributions decreases from $\sim30^\circ$ to $\sim10^\circ$ for group and cluster halos, with slightly stronger alignment for red galaxies compared to blue ones of similar mass. Simulations also indicate that the angular momentum directions of the gaseous and dark matter components are typically misaligned by $\sim30^\circ$ \citep{2010MNRAS.404.1137B}. Interestingly, the alignment of the minor axes of blue central galaxies and their halos ($\sim40^\circ$) is comparable to the typical alignment between a halo's angular momentum and its minor axis \citep{2007MNRAS.378.1531K}. 

As discussed above, observational studies generally find preferential—but not perfect—alignment between central galaxies and their host halos, with typical projected angles of $20^\circ$ to $40^\circ$ and stronger signals in red and higher-mass systems. Cosmological simulations likewise predict central-halo alignments with a similar qualitative dependence on mass and galaxy type, and often report somewhat tighter alignment for red or massive halos in 3D. Direct comparison, however, is nontrivial: observations measure projected  {spin axis} of stellar componts, whereas simulations often quantify 3D halo or stellar angular momentum or shape; projection effects and differing operational definitions can therefore introduce systematic offsets. In addition, such preferential alignment is thought to arise from the complex interplay of dynamical processes during galaxy and group formation \citep[see, e.g.,][]{2015SSRv..193....1J,2015SSRv..193...67K,2015SSRv..193..139K}.

While the major axis provides a convenient observational proxy, it does not fully capture the dynamical state of the system. A more physically motivated and dynamically relevant quantity is the angular momentum, or spin, which directly traces the cumulative effects of tidal torques, mergers, and accretion. The key question, therefore, is whether there exists a direct physical connection between the spin of the dark matter halo and that of the central galaxy. Theoretically, the BCG is expected to inherit the angular momentum of its host halo, but processes such as gas cooling, feedback, and repeated mergers may decouple the spin of the central galaxy from that of the halo \citep{2012ApJS..203...17R, 2015MNRAS.448.3391C}.

In this Letter, we use spin as a direct probe to systematically investigate whether the alignment between galaxy groups and their central galaxies, previously established using shape measurements, can also be reproduced when using spin. By applying newly developed observational techniques to large, homogeneous samples, we aim to assess the extent to which spin alignment reflects the connection between central galaxies and their host dark matter halos, and to explore its dependence on galaxy and group properties.

\begin{figure*}
\centering
\plotone{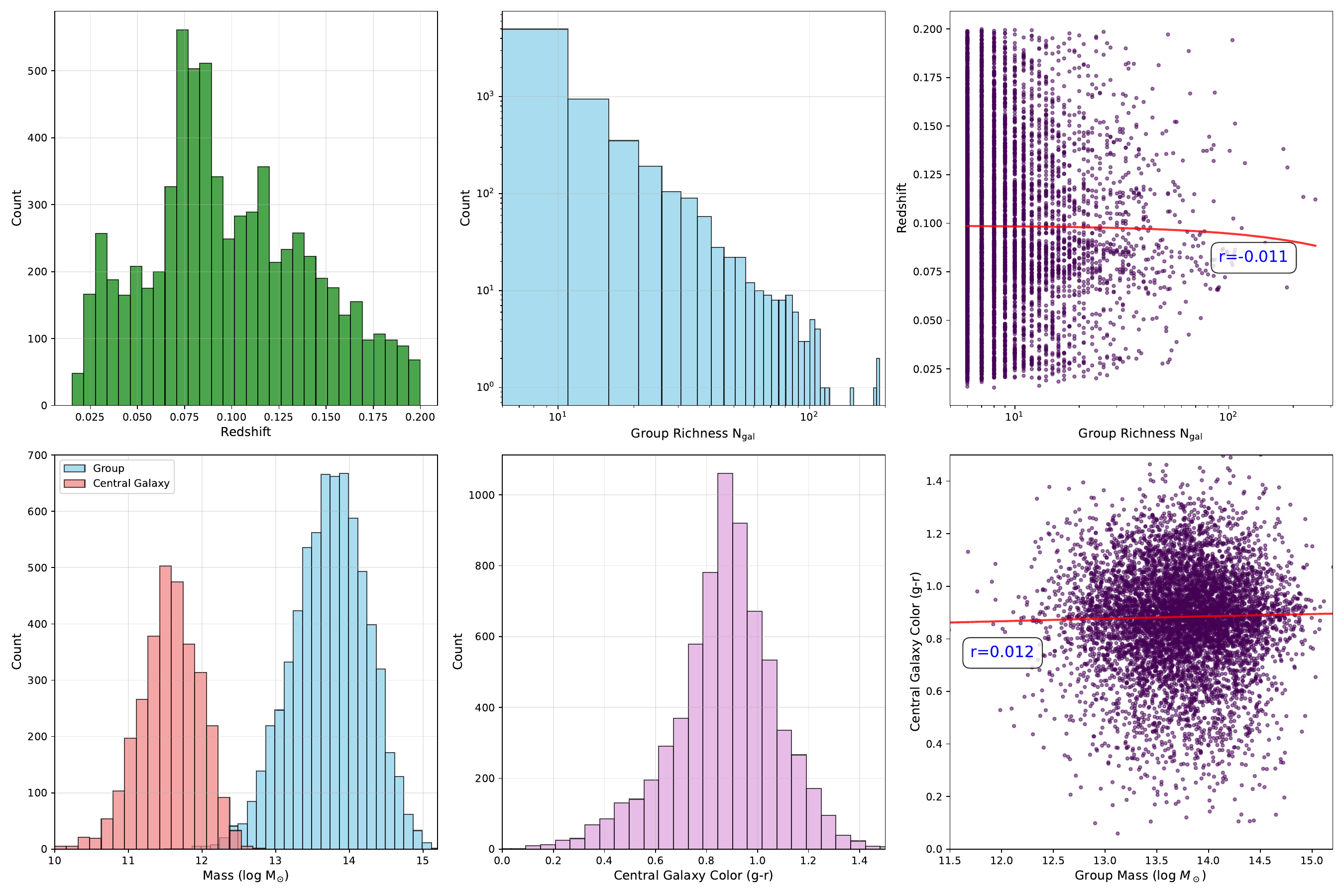}
\caption{
 {Distributions of key properties for the sample of galaxy groups and their central galaxies. 
\textbf{Top right:} Redshift distribution of galaxy groups. 
\textbf{Top middle:} Distribution of group richness ($\rm N_{gal}$). 
\textbf{Top left:} Correlation between group richness and redshift. The red line indicates the best-fit linear relation, with the correlation coefficient $r$ quantifying the strength of the correlation. 
\textbf{Bottom left:} Distribution of group masses and the stellar masses of central galaxies ($\log M_{\odot}$). 
\textbf{Bottom middle:} Distribution of central galaxy colors ($g-r$). 
\textbf{Bottom right:} Relation between the central galaxy color $(g-r)$ and the group mass $\log M_{\odot}$. The red line shows the best-fit linear relation, while the correlation coefficient $r$ reflects the strength of the linear correlation.}
}
\label{fig:f1}
\end{figure*}

\section{Data and Methodology}
\label{sec:method}

\subsection{Galaxy and Group Catalog}
\label{sec:group_sample}

Our analysis is based on a spectroscopic galaxy sample derived from the Sloan Digital Sky Survey Data Release 12 \citep[SDSS,][]{2015ApJS..219...12A}, as compiled by \citet{2017A&A...602A.100T}. To ensure the reliability of the dataset, we impose a Petrosian $r$-band magnitude cut of $m_r = 17.77$~mag, thereby excluding objects with low signal-to-noise ratios or unreliable photometry. After this selection, the sample comprises 584,449 galaxies, each associated with a group identifier, CMB-corrected redshift, and a suite of physical and observational parameters. The redshift distribution of our group sample spans $z = 0.015$ to $z = 0.2$, with a median redshift of $z_{\rm med} = 0.09$, as shown in the top-left panel of Figure~\ref{fig:f1}. 

Galaxy groups are identified using the modified \citep{2014A&A...566A...1T, 2017A&A...602A.100T} Friends-of-Friends (FoF) algorithm \citep{1985ApJ...295..368B, 2009MNRAS.399..497D}. In this approach, the transverse comoving linking length is adaptively scaled with redshift as
$d_{\rm LL}(z) = 0.34 \text{ Mpc} \times \left[1 + 1.4 \arctan\left(\frac{z}{0.09}\right)\right]$
and an anisotropic linking with a radial-to-transverse ratio $b_{\parallel}/b_{\perp} = 12$ is adopted to account for redshift-space distortions. Group membership is subsequently refined following \citet{2016A&A...588A..14T}: for FoF systems with $\rm N_{gal} \ge 7$, a multimodality analysis using \texttt{mclust} is applied to split subcomponents, followed by an iterative removal of unbound galaxies based on the virial radius ($R_{200}$) and the escape velocity inferred from the velocity dispersion ($\sigma_v$) and the sky-projected extent ($\sigma_{\rm sky}$). Small groups are also re-detected among the excluded galaxies.
Potentially merging systems are flagged when the 3D comoving separation between group centres is smaller than the sum of their characteristic radii ($R_{\rm group}$). The characteristic radius $R_{\rm group}$ is defined as a conservative combination of $R_{200}$ and the maximum observed sky-plane radius ($R_{\rm max}$) \citep{2017A&A...602A.100T}. This methodology does not impose a strict requirement for full dynamical relaxation. In the \citet{2017A&A...602A.100T} catalogue, out of 88{,}662 groups with at least two members.
 {Among them, 498 potentially merging systems are identified, involving 503 galaxy groups, as each system may consist of more than one group.}

\subsection{Group Spin Identification}
We describe two complementary methods to estimate the spin of galaxy groups.

{\bf Method 1:}
To identify the possible rotation of a galaxy group, we employ a general observational method that has been widely adopted in recent studies. This approach was first applied by \cite{2021NatAs...5..839W} to measure the spin of large-scale filamentary structures, and has since been utilized in several works, including \cite{2025ApJ...982..197T} and \citet{2025ApJ...983..100W}. More recently, \cite{2025arXiv250813597T} applied this method to quantify the coherent spin of galaxy clusters. We measure cluster spin with a side-to-side redshift contrast method. A trial dividing line (with angle $\theta$  to a given reference line ) through the cluster center is rotated from 0$^\circ$ to 180$^\circ$, splitting members into two hemispheres (each with $\geq$3 galaxies, $\geq$6 in total). For each angle $\theta$ we compute the mean redshifts of the two sides and take the absolute difference; the maximum over angles, $\Delta Z_{\rm max}$, is the spin-amplitude proxy, and the corresponding angle $\theta_{\rm max}$ defines the projected rotation axis.  This non-parametric method yields a well-defined axis for alignment studies, but its slightly sensitivity depends on member richness and spatial coverage. We refer readers to Method and Appendix section in \cite{2025arXiv250813597T} for more details.

{\bf Method 2:}
As a complement to Method~1, we employ a second approach. Following \cite{2025ApJ...983L...3R}, we estimate the group spin in the projected plane. The brightest member defines the group center and the line-of-sight unit vector $\mathbf{e}_c$. Member positions are transformed from survey coordinates to Cartesian and expressed as sky-plane offsets $\mathbf{r}_i$ relative to the center. Spectroscopic redshifts give radial velocities $v \simeq c\,z_{\rm obs}$; the relative line-of-sight velocity of the $i$-th member is
\[
\mathbf{v}_i \;=\; (v_i - v_c)\,\mathbf{e}_c \,,
\]
with $v_c$ the center's radial velocity. By construction, $\mathbf{r}_i$ lies in the sky plane and $\mathbf{v}_i$ is parallel to $\mathbf{e}_c$, so the resulting spin proxy is orthogonal to the line of sight.

The projected group spin is then
\[
\mathbf{L}_p \;=\; \sum_{i=1}^{N} \,\big(\mathbf{r}_i \times \mathbf{v}_i\big)\,,
\qquad
\mathbf{e}_{L_p} \;=\; \frac{\mathbf{L}_p}{\lVert \mathbf{L}_p \rVert}\,,
\]
where $\mathbf{e}_{L_p}$ denotes the projected spin axis. 

 {To ensure robustness and consistency, we first apply the Method~1 selection criterion, which requires groups to contain at least six galaxies so that each hemisphere has at least three members for a stable spin-axis determination. This yields a parent sample of 6,873 groups. Method~2 is then applied to the same set of groups, so that the two methods are based on an identical underlying sample and the differences in the results arise solely from the methodological approach.}

 {The distribution of group richness ($\rm N_{gal}$) is shown in the histogram in the top-middle panel of Figure~\ref{fig:f1}. Most groups are poor systems with relatively small membership, and the number of rich systems decreases rapidly with increasing $\rm N_{gal}$. This highly skewed distribution, with the majority of groups having $\rm N_{gal}<10$ (4,588 groups, $\sim$67\% of the total). We also examined the correlation between group richness ($\rm N_{gal}$) and redshift in the top-right panel. The fitted trend yields a correlation coefficient of $r=-0.011$, indicating that there is no statistically significant dependence.}
The mass distribution of these selected groups is shown in the  {bottom-left} panel of Figure~\ref{fig:f1},  {represented by the sky-blue histogram, with a mass range of approximately $12$--$15$, in the units of $\rm log \ M_{\odot}$.}

We assess the consistency between the two group spin estimators by computing the projected angular separation between their spin axes. For the 6,873 groups with $N_{\rm sat} \ge 6$, the distribution of $\Delta\psi$ has a median of $24.6^\circ$, a mean of $30.5^\circ$, and a 68\% interval of [$6.5^\circ$, $58.2^\circ$]. The fractions with $\Delta\psi < 15^\circ$ and $\Delta\psi < 30^\circ$ are 34.5\% and 57.7\%, respectively. Using axial statistics (doubling the angles), the Rayleigh test strongly rejects isotropy ($p \approx 0.0e+00$; i.e., $p \ll 10^{-300}$), and a von Mises fit yields a concentration parameter of $\kappa = 2.01 \pm 0.03$. These results demonstrate that the two spin estimators are highly consistent in their inferred directions.

 {The above statistical comparison quantitatively demonstrates the overall consistency between the two estimators. Building on this, we next compare their relative performance under different sampling conditions (sparse versus isotropic), as summarized below.}
Despite a certain difference in the specific directions estimated by the two spin measurement methods (with a median angular separation of approximately 25$^\circ$), statistical tests show that both significantly deviate from a random distribution and exhibit a highly consistent overall trend. Method~1 is a non-parametric LOS-based axis finder: for each trial angle $\theta$ it bisects the members, maximizes the side-to-side mean redshift contrast $\Delta Z(\theta)$, and adopts $\theta_{\rm max}$ as the projected rotation axis (equal weights; dipolar LOS pattern, no rotation model). Method~2 is a vector angular-momentum proxy, $\mathbf{L}_p$, coupling sky-plane positions with LOS velocities; it up-weights large $|\mathbf{r}_i|$ and measures net projected specific angular momentum. Practically, Method~1 is more robust for sparse or azimuthally incomplete sampling; Method~2 performs best with isotropic spatial coverage and broad radial extent but is more sensitive to asymmetries and distant interlopers. The methods are complementary: Method~1 detects LOS velocity dipoles, Method~2 captures spatially weighted spin.


\subsection{Central Galaxy Spin}
\label{subsec:Spin_central}

Within each group, the central galaxy is typically the brightest member, although adopting the most massive member as the central yields consistent results. This choice does not affect our conclusions. The stellar masses of central galaxies, $M_\star$, are estimated from the $r$-band absolute magnitudes and ($g-r$) colors using the mass-to-light ratio equation $\rm log(M_\star/L_r)=1.097(g-r)-0.306$ \citep{2003ApJS..149..289B}. The distribution of central galaxy stellar masses is shown in the   {bottom-left} panel of Figure~\ref{fig:f1},  {represented by the light-red histogram.} The color distribution of central galaxies, as measured by their $g-r$ values, is shown in the  {bottom-middle} panel.  {In the bottom-right panel, we check the the relation between group mass ($\log M_\odot$) and the central galaxy color ($g-r$). The wide scatter and a small correlation coefficient of $r=0.012$ indicate that group mass and central color are essentially uncorrelated.}
The morphological classification of central galaxies is based on the probabilities provided in the \citet{2017A&A...602A.100T}, which assigns each galaxy a probability of belonging to one of four types: ``h\_e'', ``h\_s0'', ``h\_sab'', and ``h\_scd''. We classify a central galaxy as elliptical if $\rm h\_e > (h\_s0 + h\_sab + h\_scd)$, and as spiral if $\rm h\_e < (h\_s0 + h\_sab + h\_scd)$. Using this criterion, we identify 2,993 elliptical and 3,880 spiral central galaxies in our sample.

The spin vector of each central galaxy is determined following the methods of \citet{2012ApJ...744...82V, 2021MNRAS.504.4626K, 2015ApJ...798...17Z}. Specifically, the spin  {axis} components are calculated as
\[
\begin{aligned}
S_x &= \cos \alpha \cos \delta \sin \zeta + \cos \zeta (\sin \phi \cos \alpha \sin \delta - \cos \phi \sin \alpha), \\
S_y &= \sin \alpha \cos \delta \sin \zeta + \cos \zeta (\sin \phi \sin \alpha \sin \delta + \cos \phi \cos \alpha), \\
S_z &= \sin \delta \sin \zeta - \cos \zeta \sin \phi \cos \delta,
\end{aligned}
\]
where $\alpha$ and $\delta$ denote the right ascension and declination, respectively, and $\phi$ is the position angle. 
Note that, because our galaxy sample lacks IFU kinematics, we cannot determine the sense of rotation; the above formula therefore yields only the spin axis (an axial quantity). For directed-spin measurements, we refer readers to an study of a MaNGA \citep{2017AJ....154...86W, 2015ApJ...798....7B} subsample, which inferred the IFU-based  spin direction and, for the first time observationally, reported a tendency for anti-parallel alignments \citep{2025ApJ...987L..30W} between low-mass spiral galaxies and filament spins.

\begin{figure*}
\centering
\plotone{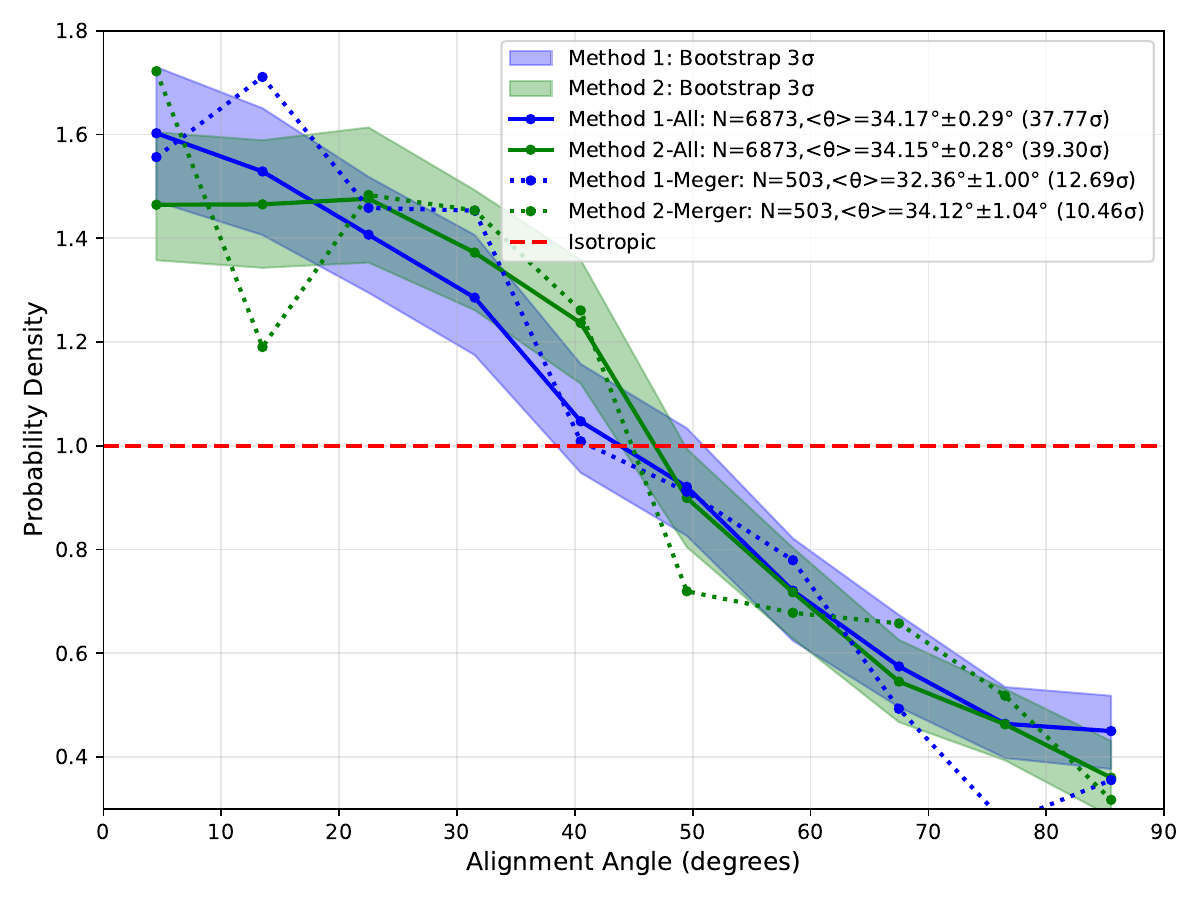}
\caption{Probability density of the projected alignment angle between the spin of galaxy groups and that of their central galaxies. The  {solid} curves labeled  {``Method 1-All'' and ``Method 2-All'' } show results  {for all groups}. 
 {The dotted curves labeled ``Method 1-Merger'' and ``Method 2-Merger'' show the corresponding results for merging groups. For clarity, only the 99.73\% ($3\sigma$) bootstrap confidence intervals for all sample are shown.}
The ``Isotropic'' curve denotes the expectation for random  {spin axes} in projection. 
 {In the legend, we provide the sample size $N$, the mean alignment angle $\langle\theta\rangle$ with its corresponding uncertainty, and, in parentheses, the detection significance relative to isotropy.}
}
\label{fig:f2}
\end{figure*}

\subsection{Alignment Angle Calculation}
\label{subsec:Alignment_angle}

With the spin  {axes} of both the central galaxies and their host groups defined, we quantify  {their relative alignment by calculating the projected angle between the two axes}. For each group--central galaxy pair, the projected angle is determined from the dot product of their respective projected spin   {axis}.  {It is important to note that, due to the limitations of our observational data and the absence of IFU-based kinematic measurements, we are only able to recover the \textit{projected spin axes} of galaxies and groups, not their true three-dimensional spin \textbf{directions}.}  {Accordingly, the alignment angle is defined strictly in projection and restricted to the range $0^{\circ}$--$90^{\circ}$, where $0^{\circ}$ indicates perfect parallel alignment and $90^{\circ}$ indicates perfect perpendicular alignment in the projected plane.} A random (isotropic) distribution between the spins of central galaxies and groups would result in a mean projected angle of $45^{\circ}$.  {A mean value below $45^{\circ}$ then implies a statistical preference for parallel alignment, while a value above $45^{\circ}$ indicates a preference for perpendicular alignment.}

\begin{figure*}
\centering
\plotone{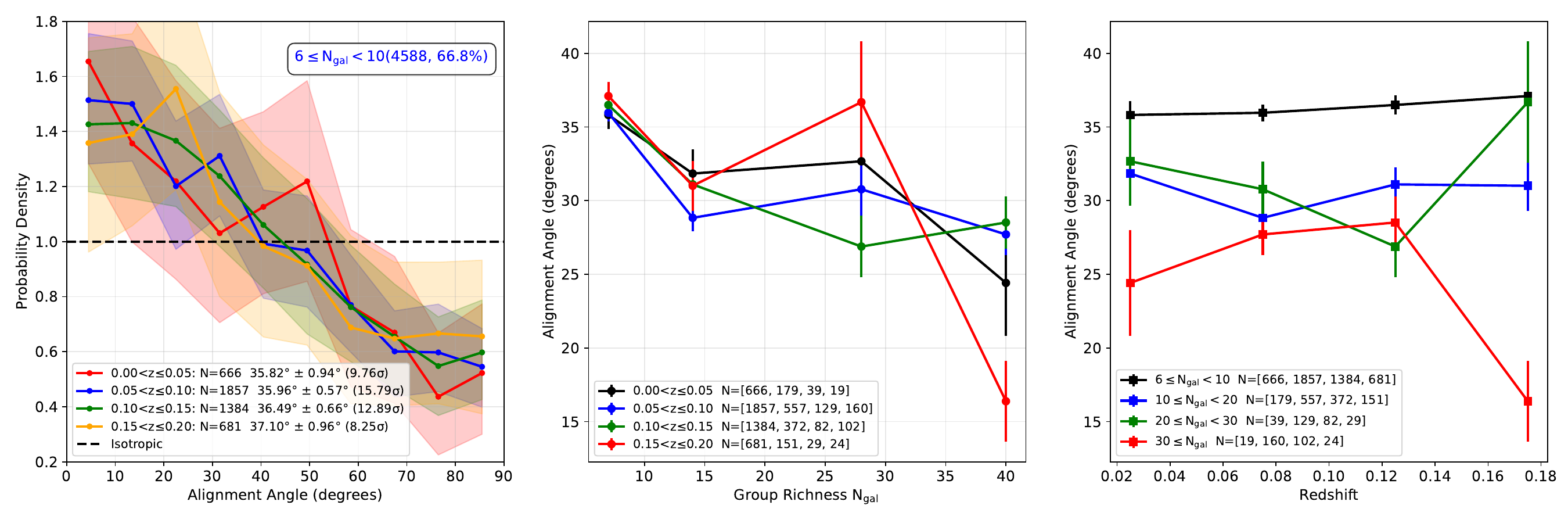}
\caption{
 {
\textbf{Left:} Same as Fig.~\ref{fig:f2}, but for groups with $6 \leq  \rm N_{gal} <10$ in different redshift bins. 
Shaded regions indicate the $3\sigma$ confidence intervals. 
The legend lists the redshift range, sample size, mean alignment angle with its $1\sigma$ error, and the detection significance relative to random samples. 
The black dashed line marks the isotropic expectation. 
\textbf{Middle:} Alignment angle as a function of group richness $\rm N_{gal}$ for several redshift intervals, with error bars showing the $1\sigma$ uncertainties.In the legend, $N$ gives the sample sizes for the four richness bins shown by the points on each curve.
\textbf{Right:} Alignment angle as a function of redshift for different richness bins $\rm N_{gal}$, with $1\sigma$ error bars. $N$ gives the sample sizes for the four redshift bins shown by the points on each curve.
}}
\label{fig:f3}
\end{figure*}

\section{Result}
\label{sec:result}

To investigate the relationship between the spin  {axes}  of galaxy groups and the spin of their central galaxies, we analyze the projected alignment angle distributions and their dependence on key physical properties.

Figure~\ref{fig:f2} presents the probability density function (PDF) of the projected angle between the spin of each galaxy group and the spin of its central galaxy. Results from two independent estimators (``Method~1'' and ``Method~2'') are shown, with shaded bands indicating the 99.73\% ($3\sigma$) bootstrap confidence intervals; the red dashed curve denotes the isotropic (random) expectation in projection. Both methods display a pronounced excess at small angles relative to isotropy and yield consistent mean alignment angles of $34.17^\circ \pm 0.29^\circ$ and $34.15^\circ \pm 0.28^\circ$, respectively, substantially below the isotropic expectation of $45^\circ$. The deviations correspond to detection significances of $37.77\sigma$ and $39.30\sigma$, providing extremely strong statistical evidence for a preferential alignment between group and central spins. These results indicate that the angular momentum of the central galaxy is closely coupled to that of its host group. In the analyses that follow, we adopt Method~1 as our fiducial estimator.

 {In addition to the full sample, we also analyze the subsample of 503 merging groups (comprising 7.3\% of the total, shown by the dotted curves in Figure~\ref{fig:f2}). The mean alignment angles are $32.36^\circ \pm 1.00^\circ$ (Method~1-Merger) and $34.12^\circ \pm 1.04^\circ$ (Method~2-Merger), corresponding to detection significances of $12.69\sigma$ and $10.46\sigma$, respectively. The reduced sample size leads to larger uncertainties, Method~1 appears to indicate a slightly stronger alignment signal, based on the mean alignment angle, in the merging subsample when compared to the full sample. By contrast, Method~2 yields nearly identical results for the merging and full samples ($34.12^\circ$ vs. $34.15^\circ$), demonstrating its higher stability. Overall, the merging group results generally fall within the $3\sigma$ confidence interval of the full sample, and the mean angles differ only marginally from those of the full sample, further reinforcing the robustness of the alignment signal. However, given the limited number of merging systems, the current data do not allow us to definitively distinguish any significant differences in alignment angles between merging and non-merging groups.}

\begin{figure*}
\centering
\plotone{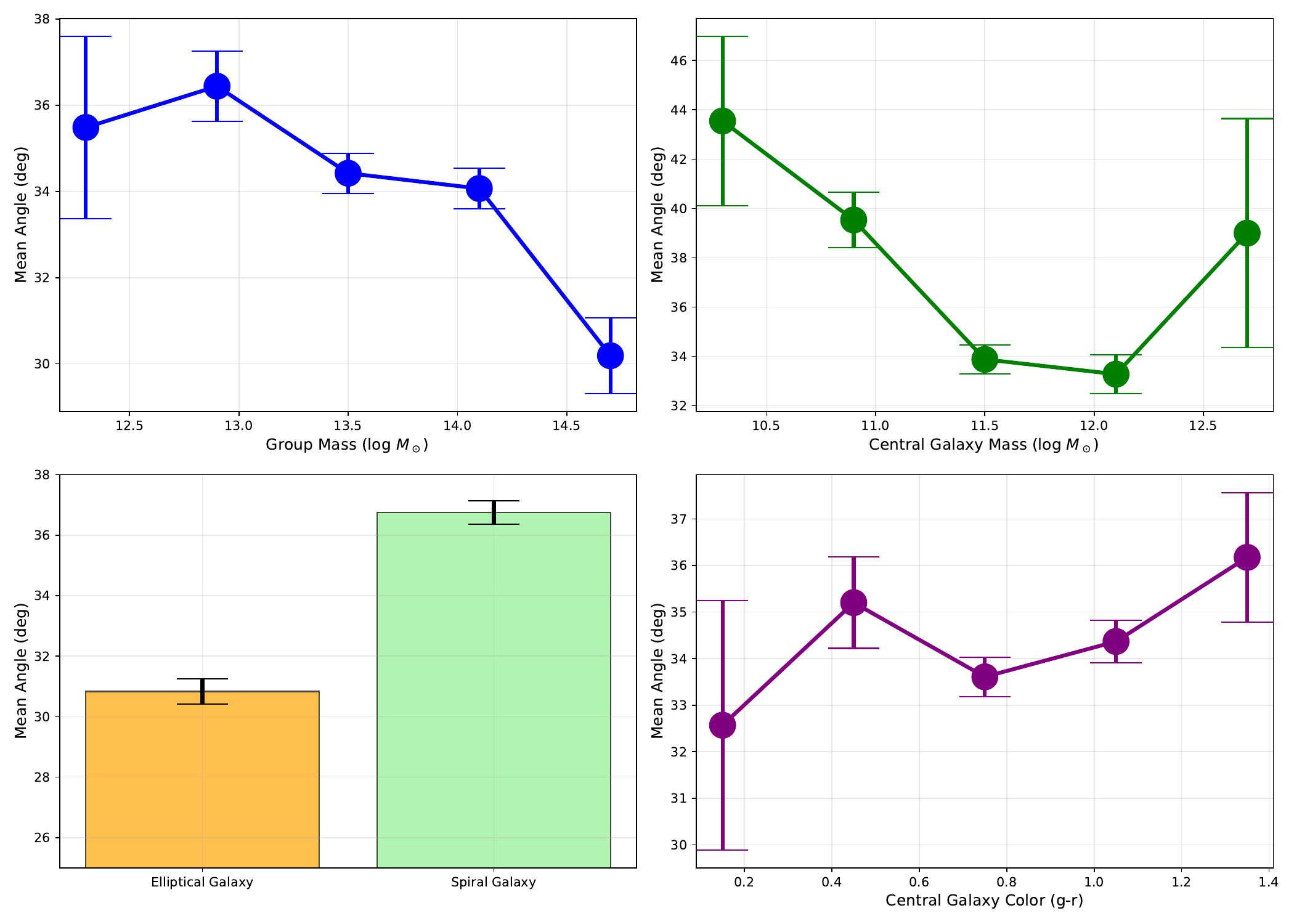}
\caption{
Dependence of the mean projected alignment angle between group spin and central galaxy spin on various physical properties. 
\textbf{Top left:} Mean angle as a function of group mass (log $M_\odot$). 
\textbf{Top right:} Mean angle as a function of central galaxy stellar mass (log $M_\odot$). The red dashed line in each panel indicates the expected value for random (isotropic).
\textbf{Bottom left:} Mean angle for elliptical and spiral central galaxies.
\textbf{Bottom right:}  Mean angle as a function of central galaxy color ($g-r$). 
All error bars represent the standard error of the mean in each bin or group. 
}
\label{fig:f4}
\end{figure*}

In Table~\ref{tab:align_by_richness}, the alignment signal is detected at high significance across all richness cuts. For the full sample with $\rm N_{gal}\!\ge\!6$ (6873 groups), the measured amplitude is $34.17\!\pm\!0.29$ (37.77$\sigma$), which also listed in the legend in the  {Figure~\ref{fig:f2}}. As the richness threshold increases, the sample size decreases and the uncertainties grow, but the detections remain robust. The amplitude exhibits a slight strengthening with increasing richness (e.g., from $34.17$ at $\rm N_{gal}\!\ge\!6$ to $26.90$ at $\rm N_{gal}\!\ge\!30$ across progressively richer subsamples), consistent with expectations that systems with more members provide more stable  {axis}  estimates and reduced shot noise. 

\begin{table}[t]
\centering
\caption{Alignment signal versus group richness. For each richness threshold, we list the number of groups included and the measured alignment signal (value $\pm$ uncertainty), with the corresponding detection significance in parentheses.}
\label{tab:align_by_richness}
\begin{tabular}{l|cc}
\hline
Richness bin & Number of groups & Signal  \\
\hline
$\rm N_{gal} \ge 6$    & 6873 & $34.17 \pm 0.29 (37.77\sigma$)  \\
$\rm N_{gal} \ge 10$  & 2285 & $29.96 \pm 0.45 (33.73\sigma$)  \\
$\rm N_{gal} \ge 20$  & 642   & $28.17 \pm 0.75 (22.37\sigma$)  \\
$\rm N_{gal} \ge 30$  & 322   & $26.90 \pm 0.99 (18.35\sigma$)  \\
\hline
\end{tabular}
\end{table}

 {To assess possible biases related to sample selection, we further examined the distribution of alignment angles as a joint function of group richness ($\rm N_{gal}$) and redshift. The results are presented in Figure~\ref{fig:f3}, where the left panel shows the probability distribution of alignment angles in several redshift bins, the middle panel illustrates the variation of the mean alignment angle with richness in different redshift bins, and the right panel displays its redshift dependence for different richness ranges in different richness bins. }

 {As shown in the left panel, this subsample corresponds to groups with $6 \leq N_{\rm gal} < 10$, which make up the majority of the catalog ($\sim 67\%$, or 4,588 systems). Across all redshift intervals, the observed distributions deviate significantly from isotropy, consistently favoring small-angle alignments. From the mean angles listed in the legend, there appears to be a mild decrease of the average value with increasing redshift, suggesting a slightly stronger alignment signal at higher $z$. However, when considering the full distributions, all curves lie well within their respective $3\sigma$ confidence ranges, indicating that the differences between redshift bins are not statistically significant.}

 {The middle panel presents the relation between the mean alignment angle and group richness $N_{\rm gal}$, with different curves corresponding to redshift intervals. Overall, the alignment signal strength increases systematically with $N_{\rm gal}$, consistent with the trend reported in Table~1. For poorer groups ($N_{\rm gal} \lesssim 30$), the curves corresponding to different redshift bins are largely consistent within the $1\sigma$ uncertainties, showing little redshift dependence. In contrast, for richer groups the differences become more pronounced, with the highest-redshift subsample ($0.15 \le z < 0.20$, red curve) exhibiting an apparently stronger alignment (smaller mean angle) but low sample size.}

 {The right panel presents the redshift evolution of the mean alignment angle for different richness bins. The lowest-richness groups ($6 \le N_{\rm gal} < 10$, black curve) remain nearly constant at $\sim 35^{\circ}$–$37^{\circ}$ across the full redshift range, showing no clear evolution and characterized by small uncertainties owing to their large sample size. Similarly, the $10 \le N_{\rm gal} < 20$ bin (blue curve) exhibits a nearly flat trend around $\sim 31^{\circ}$, with only mild fluctuations that are well within the $1\sigma$ uncertainties. Groups with $20 \le N_{\rm gal} < 30$ (green curve) show somewhat larger variations: their alignment angle decreases to $\sim 27^{\circ}$ at intermediate redshift ($z \sim 0.12$) but rises again toward high redshift, though the differences are not statistically significant given the error bars. In contrast, the richest systems ($N_{\rm gal} \ge 30$, red curve) appear to decline toward higher redshift, reaching values as low as $\sim 15^{\circ}$ in the last bin.}

 {These results suggest that the observed alignment signal is not an artifact of sample selection, but rather an intrinsic property of galaxy groups that depends on richness and on cosmic epoch. Low-richness systems show little evolution with redshift, consistent with their more relaxed dynamical state, whereas rich systems reveal enhanced alignment at earlier times, plausibly linked to their ongoing dynamical assembly within large-scale structures. As indicated by Table~1, however, groups with $N_{\rm gal}\ge 30$ constitute only $\sim 5\%$ of the total sample (322 systems), and once further divided into redshift bins the numbers in each subsample become very small (see the legend in the Figure~\ref{fig:f3}). Consequently, the apparent stronger alignment at high redshift for the richest groups, including the sharp drop seen in the right panel, should be treated with caution, as it may largely reflect small-number statistics rather than a robust evolutionary feature.}

The dependence of the mean projected alignment angle on various physical properties is shown in Figure~\ref{fig:f4}. In the top left panel, the mean alignment angle is plotted as a function of group mass (log~$M_\odot$). Across the mass range from approximately 12.5 to 14.5 in log~$M_\odot$, the mean angle remains consistently below the isotropic expectation of $45^\circ$, with values ranging from about $30^\circ$ to $35^\circ$. There is a slight trend of decreasing mean angle with increasing group mass, suggesting that more massive groups may exhibit marginally stronger spin alignment with their central galaxies.

 {The top right panel shows the mean alignment angle as a function of central galaxy stellar mass ($\log M_\odot$). A clear mass dependence is evident: at lower stellar masses ($\log M_\ast \lesssim 11$), the mean alignment angle is relatively large ($\sim 42^\circ$--$44^\circ$), which close the $45^\circ$ expected from the random distribution, while with increasing stellar mass the angle systematically decreases, reaching a minimum of $\sim 33^\circ$ around $\log M_\ast \sim 12$. At the highest-mass end ($\log M_\ast \gtrsim 12.5$), the angle shows a slight upturn but with large uncertainties due to the small sample size. This overall trend indicates that more massive central galaxies tend to have their spins more closely aligned with the angular momentum of their host groups, implying a stronger coupling between the galaxy and group angular momentum in high-mass systems.}


The bottom left panel compares the mean alignment angle for elliptical and spiral central galaxies. Elliptical galaxies show a mean angle of about $31^\circ$, while spiral galaxies have a higher mean angle of approximately $37^\circ$. Both values are significantly below $45^\circ$, but the alignment is notably stronger for ellipticals, implying that morphology plays a role in the degree of spin alignment.

The bottom right panel presents the mean alignment angle as a function of central galaxy color ($g-r$), covering a color range from about 0.2 (bluer galaxies) to 1.4 (redder galaxies). In this context, a smaller $g-r$ value corresponds to bluer, more star-forming central galaxies, while a larger $g-r$ value indicates redder, more quiescent systems. The results show a certain dependence of the alignment angle on color: as the $g-r$ value increases, the mean alignment angle also increases, approaching the isotropic value of $45^\circ$. This trend implies that the spin alignment between groups and their central galaxies becomes weaker for redder central galaxies. In other words, bluer central galaxies tend to have a stronger spin alignment with their host groups, while redder galaxies show a weaker alignment. This dependence suggests that the recent star formation history or stellar population age, as traced by the $g-r$ color, may play a role in modulating the degree of spin alignment.

 {Figure~\ref{fig:f4} (top-left panel) demonstrates a clear mass dependence, with more massive groups exhibiting stronger alignment. One might therefore expect that this trend should be primarily driven by red central galaxies, since they are generally more prevalent in high-mass groups. However, the bottom-right panel of Figure~\ref{fig:f4} shows that the alignment signal appears slightly stronger for blue centrals, which seems counterintuitive at first glance. This apparent discrepancy can be understood in light of Figure~\ref{fig:f1} (bottom-right), which shows that the correlation between central galaxy color and group mass in our sample is very weak. As a result, the mass and color dependencies act largely independently. The enhanced alignment in massive groups reflects the stronger tidal coherence of their host halos, while the weak color effect is not directly tied to group mass.  The stronger alignment observed in blue galaxies likely arises from their better preservation of primordial spin orientations compared to their red counterparts, which may arise because they have not yet undergone significant dynamical processing within their groups, so the influence of group mass is not fully imprinted. Such conjectures clearly warrant further investigation through dedicated cosmological simulations.}

\section{Summary and Discussion}\label{sec:sum_dis}

In this Letter, we use a spectroscopic galaxy sample from the SDSS to systematically investigate the alignment between the spin of galaxy groups and that of their central galaxies. By applying a novel spin estimator for galaxy groups and analyzing how the alignment signal depends on various physical properties, we provide new observational insights into spin alignment in the group–central galaxy system. Our main findings can be summarized as follows:

\begin{itemize}
    \item We find a statistically significant alignment between the spin of galaxy groups and the spin of their central galaxies, with a mean projected alignment angle of $34.17^\circ \pm 0.29^\circ$, which is much lower than the isotropic expectation of $45^\circ$. The deviation from randomness corresponds to a confidence level of $37.77\sigma$.
    
    \item This alignment persists across a wide range of group and galaxy properties, but shows clear dependencies:
    \begin{itemize}
        \item The alignment is generally stronger in more massive groups.
        \item  {The alignment generally becomes stronger with increasing stellar mass.}
        \item Morphology matters: elliptical central galaxies show a stronger alignment with their host groups than spiral galaxies.
        \item Color dependence is also observed: bluer central galaxies exhibit stronger spin alignment with their groups, while redder galaxies show weaker alignment.
    \end{itemize}
    
\end{itemize}


Our results provide new observational evidence for a statistically significant alignment between the spin of galaxy clusters and that of their central galaxies. This finding adds a dynamical perspective to the long-standing question of how central galaxies are connected to their host clusters and dark matter halos.

Previous studies have extensively explored the alignment between galaxies and their host halos primarily from the perspective of shapes and major axes \citep{2015SSRv..193....1J,2015SSRv..193...67K,2015SSRv..193..139K}. Observational and simulation-based works have reported typical alignment angles in the range of $20^\circ\sim40^\circ$ \citep{2004MNRAS.347..895H, 2008MNRAS.385.1511W, 2009ApJ...694..214O, 2014MNRAS.441..470T, 2015MNRAS.453..469T}, with variations depending on galaxy color, mass, and environment. Our results, based on spin measurements, reveal a statistically significant alignment between the spin of galaxy clusters and their central galaxies, with a mean projected angle of $34.17^\circ \pm 0.29^\circ$. This value is broadly consistent with the lower end of the alignment range found in previous shape-based studies, suggesting that spin can serve as a robust dynamical tracer of the connection between central galaxies and their host clusters. The observed dependencies on mass, morphology, and color further support the idea that both baryonic and dynamical processes play important roles in shaping the alignment signal.

Our study extends these investigations by focusing on the spin  {axis} as a direct, dynamically motivated tracer of alignment. To our knowledge, this is the first observational study to systematically probe the alignment between galaxy clusters and their central galaxies from a dynamical perspective, using spin as a direct tracer, complementing previous shape-based analyses \citep{2015SSRv..193....1J,2015SSRv..193...67K}. We find that the spin alignment between clusters and their central galaxies is not only statistically significant, but also exhibits clear dependencies on mass, morphology, and color. The stronger alignment observed for elliptical and bluer central galaxies suggests that both morphology and recent star formation history play a role in modulating the coupling between central galaxies and their host clusters.

These results suggest that the spin of central galaxies is closely linked to that of their host clusters, and that this connection is modulated by mass, morphology, and star formation history. Our findings provide new insights into the co-evolution of galaxies and clusters, and demonstrate the value of spin as a physically motivated probe of galaxy--cluster connections.

 {The robustness of our measurements may in principle be affected by several observational and methodological uncertainties. One possible concern is a dependence on redshift. Since our sample is limited to $z < 0.2$, strong evolutionary effects are not expected, yet subtle trends could still bias the results. To test this, we examined the alignment signal in joint bins of group richness and redshift (Figure~3). Across all redshift slices, the alignment distributions deviate significantly from isotropy, indicating that the observed signals are not induced by redshift selection alone. For low-richness groups ($6 \leq N_{\rm gal} < 10$), which account for $\sim 67\%$ of the catalog (4,588 systems), the mean alignment angle remains nearly constant at $\sim 35^\circ$--$37^\circ$ over the full redshift range, showing no clear evidence for evolution. Groups with $10 \leq N_{\rm gal} < 20$ exhibit a similarly stable signal around $\sim 31^\circ$. In contrast, the richest systems ($N_{\rm gal} \geq 30$) appear to show stronger alignment at higher redshift, with the mean angle declining to $\sim 15^\circ$ in the highest-$z$ bin. However, these rich systems constitute only about $5\%$ of the total sample (322 systems), and their numbers become very small when further divided by redshift, so the apparent trend may largely reflect limited statistics.}

 {These results suggest that the observed alignment is not a selection artifact, but instead depends on both richness and dynamical maturity. Poor groups retain weak but persistent alignment, consistent with being more relaxed and dynamically evolved, whereas rich systems indicate enhanced alignment at earlier cosmic times, plausibly connected to anisotropic accretion along large-scale filaments and ongoing assembly processes. This interpretation is consistent with theoretical expectations in which dynamically younger and more massive halos at higher redshift preserve a stronger memory of large-scale tidal torques. A more definitive understanding of the redshift and richness dependence of spin alignment will require future detailed investigations, combining high-resolution cosmological simulations with observational samples that extend to higher redshift.}

 {Other potential sources of systematic uncertainty are less significant. The dynamical state of groups may introduce some impact, but only about $7\%$ of our sample are flagged as merging systems, insufficient to alter the global statistics. As illustrated in Figure~2, their alignment signal generally falls within the $3\sigma$ range of the overall sample. Likewise, adopting either the most massive or the most luminous member as the central makes negligible difference to the measurements. Projection effects and sparse sampling inevitably introduce scatter, particularly in low-richness groups, but the consistency of our results across two independent spin estimators and across multiple subsamples strongly supports the robustness of the detected alignment.
}

Beyond these sample-selection issues, further limitations arise from the measurement itself. Spin estimates based on projected galaxy positions and redshifts are inevitably affected by projection effects and sampling noise, especially in low-richness systems where only a small number of members are available. Interpreting the observed alignments is additionally complicated by the interplay between baryonic processes (e.g., feedback, star formation) and dynamical mechanisms (e.g., mergers, tidal torques) during the assembly of galaxies and clusters.

Overall, these considerations indicate that while our results are not dominated by selection effects, some systematic uncertainties remain, particularly in low-richness or dynamically young clusters. Future progress will rely on larger and more homogeneous spectroscopic datasets, improved spin estimators, and systematic comparisons with high-resolution cosmological simulations, which together will allow a more definitive assessment of the physical origin of the observed spin alignments.

\begin{acknowledgments}
PW acknowledge the financial support from the NSFC (No.12473009), and also sponsored by Shanghai Rising-Star Program (No.24QA2711100). This work is supported by the China Manned Space Program with grant no. CMS-CSST-2025-A03
\end{acknowledgments}




\bibliography{main}{}
\bibliographystyle{aasjournal}


\end{CJK*}
\end{document}